\documentclass[journal]{IEEEtran} 
\IEEEoverridecommandlockouts
\usepackage{cite} %
\usepackage{amsmath,amssymb,amsfonts}
\usepackage{algorithmic}
\usepackage{graphicx}
\usepackage{textcomp}
\usepackage{xcolor}
\usepackage{physics}
\usepackage{subcaption}
\usepackage{url}

\usepackage{upgreek}
\newcommand{\micron}{\ensuremath{\upmu\mathrm{m}}}

\begin{document}

\title{Transient Multiscale Workflow for Thermal Analysis of 3DHI Chip Stack
\thanks{This work is supported by funding provided by the Center for Future Energy Systems (CFES) and NYSTAR.}}

\author{
    \IEEEauthorblockN{Mohammad Elahi\IEEEauthorrefmark{1}}
    , \IEEEauthorblockN{Max O. Bloomfield\IEEEauthorrefmark{1}\IEEEauthorrefmark{2}}
    , \IEEEauthorblockN{Theodorian Borca-Tasciuc\IEEEauthorrefmark{3}}
    , \IEEEauthorblockN{Jacob S. Merson\IEEEauthorrefmark{1}\IEEEauthorrefmark{3}}
    \\
    \IEEEauthorblockA{\IEEEauthorrefmark{1}\textit{Scientific Computation Research Center, Rensselaer Polytechnic Institute, Troy, NY, USA}}\\
    \IEEEauthorblockA{\IEEEauthorrefmark{2}\textit{Center for Materials, Devices and Integrated Systems, Rensselaer Polytechnic Institute, Troy, NY, USA}}\\
    \IEEEauthorblockA{\IEEEauthorrefmark{3}\textit{Department of Mechanical, Aerospace, and Nuclear Engineering, Rensselaer Polytechnic Institute, Troy, NY, USA}}\\
    \IEEEauthorblockA{Email: mersoj2@rpi.edu; ORCID: 0000-0002-6813-6532}
}

\maketitle

\begin{abstract}
Modern package designs make use of technologies such as backside power delivery (BSPD) and 3D stacked chiplets that require accounting for the heterogeneity in back end of the line (BEOL) structures in hot-spot prediction. Multiscale homogenization strategies have been demonstrated to be effective for steady-state simulations, however accurate 3D transient simulations that include BEOL structures remain an open challenge.

In this work, we demonstrate a transient thermal workflow that accounts for the 3D heterogeneous structures in the BEOL for problems with strong- and weak- temporal scale separation under the assumption of temperature independent constitutive properties. Our workflow, based on Bloomfield et. al. 2025, automatically extracts, meshes, and homogenizes thermal properties from GDSII and OASIS files to construct thermal property maps.

Property maps (heat capacity and conductivity) have been generated for a 1~mm$\times$1~mm SoC-style model die that was constructed with LibreLane for 100$\times$100 grids with 5~\micron$\times$5~\micron{} representative volume elements (RVEs), and 50$\times$50 grids with 10~\micron$\times$10~\micron{} RVEs. The expressions for a transient effective conductivity are provided and a demonstration of the impact of the transient effects are provided for a single RVE. Finally, transient conductivity maps have been provided for a time integration timestep of \(\Delta t=0.001\).
\end{abstract}


\section{Introduction}
Thermal design of 2.5D and 3D integrated packages span length scales exceeding nine orders of magnitude, making fully resolved simulations impractical even on leadership-class supercomputers. The transient and heterogeneous nature of modern chip stacks require new methods that can resolve transients and incorporate microstructural heterogeneity and be fast enough for engineering design. A hierarchical multiscale finite-element scheme demonstrated on steady-state problems \cite{bloomfieldMultiscaleWorkflowThermal2025} can bridge these scales by coupling package-level thermal models to sub-scale analyses of back-end-of-line (BEOL) metallization. However, the acceleration of transient multiscale simulations remains an area of need. In this work, we present a transient multiscale finite-element workflow that (i) performs variationally consistent transient homogenization without assuming 1D heat flow, (ii) explicitly resolves BEOL thermal inertia, and (iii) operates on geometry constructed directly from GDSII/OASIS layouts.

The demonstrated workflow makes use of an upscaling multiscale scheme that relies upon spatial separation of scales. However, due to the wide range of materials and fast thermal loading rates, strong separation in time cannot be routinely assumed. This lack of temporal scale separation requires that microstructural models must be integrated in time.

Multiscale modeling strategies have been widely deployed to investigate package-level thermal solutions. For example, \cite{chowdhuryFastAccurateMachine2025, liuPhysicalFeaturebasedMachine2026} apply machine learning to approximate homogenized BEOL conductivities. A direct FEM-based homogenization strategy was demonstrated in \cite{bloomfieldMultiscaleWorkflowThermal2025} to extract BEOL conductivities heterogeneously integrated packages and compared to analytic solutions in  multi-layer films \cite{wastiEffectiveMediumApproximation2025}. In \cite{wangMultiscaleAnisotropicThermal2024} anisotropic properties were directly extracted from GDSII files based on an anisotropic volume-averaging strategy that relies upon uncoupled in-plane and out-of-plane heat transfer.

Multiscale acceleration strategies have also been employed for transient chip models. In \cite{barabadiMultiscaleTransientThermal2015}, the authors construct a multiscale reduced order model (ROM) through proper orthogonal decomposition (POD) of a flip chip ball grid array (FCBGA). The use of a POD ROM allows the re-use of solutions with different input power maps. They also make use of a package scale thermal solution with volume averaged material properties to provide boundary conditions to chip scale models which are broken into regions for silicon, C4s, etc. A BTE-FEM strategy was employed to extract BEOL properties in \cite{changThermalModelingAnalysis2025} and showed good agreement to pure BTE simulations. A modified alternating implicit finite difference method that assumes 1D heat transfer in the construction of the anisotropic volume averaged homogenized properties was demonstrated by \cite{nieEfficientTransientThermal2023}. A number of authors have made use of RC networks to accelerate simulations of smartphones \cite{doustiTherminator2Fast2021} and HI chips \cite{pfrommMFITMultiFidelityThermal2025}.

One common assumption in the construction of transient thermal models is that heat transport is effectively one-dimensional and that property homogenization may be performed independently in each coordinate direction. In contrast, following the variationally consistent transient multiscale formulations of \cite{larssonVariationallyConsistentComputational2010} and \cite{ramosExtensionHillMandel2017}, we make no assumption of one-dimensional heat flow. The key novelty of this work is the integration of a variationally consistent transient homogenization framework with microstructural models constructed directly from GDSII/OASIS design files. GDSII and OASIS are standard EDA layout file formats that represent chip designs as hierarchical
collections of 2D polygon layers (metals, vias, dielectrics, etc.) with associated metadata. This approach represents a first step toward high-fidelity transient simulation of 3DI chip stacks and provides a natural foundation for the systematic construction of transient reduced-order models suitable for inclusion in engineering design workflows.


\section{Multiscale Transient Model} \label{sec:MTM}
\subsection{Governing Equations}
In this method, we utilize the same set of governing equations for both the macroscale and microscale problem. For compactness, we present the strong and weak forms here with no distinction between the macroscale and microscale variables and function spaces.

The strong-form of the thermal continuity equation with no heat generation is given by
\begin{equation}
    \dot{\epsilon} + \nabla \cdot \vb{q} = 0
    \label{eq:strong-form}
\end{equation}
where \(\dot{\epsilon}\) is the rate of internal energy, \(\vb{q}\) is the heat flux vector.

Find \(\theta \in \mathcal{S}\) such that
\begin{equation}
    \int \delta \theta \dot{\epsilon} - \nabla \delta\theta \cdot \vb{q} \dd{\Omega}+\int_{\Gamma_h} \delta \theta \vb{q}\cdot \vb{n} \dd{\Gamma} = 0 \quad \forall \quad \delta \theta \in \mathcal{V}
    \label{eq:weak-form}
\end{equation}
where \(\mathcal{S}\) and \(\mathcal{V}\) are the standard trial and test function spaces that take care of the Dirichlet boundary conditions on \(\Gamma_g\), \(\Gamma_h\) is the part of the boundary with Neumann boundary conditions and \(\Gamma_g\) is the part of the boundary with Dirichlet boundary conditions such that \(\Gamma = \Gamma_h\cup\Gamma_g\) and \(\Gamma_h \cap \Gamma_g = 0\).

\subsection{Scale Transition Rules}
The microscale temperature field is decomposed based on a Taylor series expansion around each macroscale integration point as
\begin{equation}
    \theta = \bar{\theta} + \bar{\nabla} \bar{\theta} \cdot (\vb{x}-\bar{\vb{x}})+\tilde{\theta}
    \label{eq:thetaexpand}
\end{equation}
where, quantities with an overbar refer to macroscale quantities and those without correspond to microscale quantities. The last term \(\tilde{\theta}\) is referred to as the microstructural fluctuations in the literature and corresponds to the deviation from the smoothly varying macro-model. This expression leads to a set of downscaling rules
\begin{equation}
    \bar{\theta} = <\theta>,
    \label{eq:thetabar}
\end{equation}
and
\begin{equation}
    \bar{\nabla}\bar{\theta} = <\nabla \theta>,
    \label{eq:gradthetabar}
\end{equation}
where angle brackets are used to denote a volume average over the representative volume element, a small domain centered at the macroscale integration point. That is, \(<u> = \frac{1}{\Omega_\text{RVE}} \int_{\Omega_\text{RVE}} u \dd{\Omega}\). By applying Eq.~\eqref{eq:thetabar} and Eq.~\eqref{eq:gradthetabar} to Eq.~\eqref{eq:thetaexpand} the following constraints become evident:
\begin{equation}
    <\tilde{\theta}> = <\nabla \tilde{\theta} > = 0.
    \label{eq:fluctuationstozero}
\end{equation}
These constraints provide a set of allowable boundary conditions for the RVE. In this work, we use the homogeneous temperature boundary conditions given as
\begin{equation}
    \theta = \bar{\theta} + \bar{\nabla}\bar{\theta}\cdot (\vb{x}-\bar{\vb{x}}) \quad \text{on} \quad \Gamma_\text{RVE}
\end{equation}
where \(\Gamma_\text{RVE}\) is the surface of the RVE. Other common choices include the homogeneous flux boundary condition, or the periodic boundary condition. Although the periodic boundary conditions converge to consistent RVE properties faster, they are not appropriate in this setting since the underlying BEOL structure is not periodic.

To construct the upscaling rules, we make use of the Hill-Mandel principle that states that the virtual power increment at the macroscale is equal to the volume average of the virtual power increment of the microscale. That is,
\begin{equation}
\begin{split}
    \delta \bar{\theta} \dot{\bar{\epsilon}} - \bar{\nabla}\delta\bar{\theta} \cdot \bar{\vb{q}} &= <\delta \theta \dot{\epsilon} - \nabla\delta\theta \cdot \vb{q}>\\ &= 
    < (\delta \bar{\theta} + \delta \bar{\nabla} \bar{\theta} \cdot (\vb{x}-\bar{\vb{x}})+\delta \tilde{\theta} )
    \dot{\epsilon} \\&-
   (\bar{\nabla}\delta\bar{\theta}+\nabla \delta \tilde{\theta} 
    ) \cdot \vb{q}>
\end{split}
\end{equation}
Grouping terms and pulling out constants from the volume averaging
\begin{equation}
\begin{split}
    \delta \bar{\theta} \dot{\bar{\epsilon}} - \bar{\nabla}\delta\bar{\theta} \cdot \bar{\vb{q}} = &\delta \bar{\theta}<  \dot{\epsilon}> \\
                       + &\bar{\nabla} \delta \bar{\theta} \cdot< (\dot{\epsilon} (\vb{x}-\bar{\vb{x}}))-\vb{q}>\\
                       + &<\delta \tilde{\theta} \dot{\epsilon}-\nabla \delta \tilde{\theta}\cdot \vb{q}>
\end{split}
\label{eq:hill-mandel}
\end{equation}
The last term in Eq.~\eqref{eq:hill-mandel} will be zero for any valid solution of the micro problem. This can be easily seen by comparing it to the weak formulation of the microscale problem.

Finally, this leads to the identification of the following upscaling rules
\begin{equation}
    \dot{\bar{\epsilon}} = <\dot{\epsilon}>
    \label{eq:hom-energy-rate}
\end{equation}
and
\begin{equation}
\bar{\vb{q}} = <\vb{q}-\dot{\epsilon} (\vb{x}-\bar{\vb{x}})>
\label{eq:hom-heat-flux}
\end{equation}
Eq.~\ref{eq:hom-heat-flux} is a quite interesting term as it shows that the microscale RVE provides some thermal inertia to the macroscale. Also, the inclusion of \(\vb{x}-\bar{\vb{x}}\) creates a size effect that indicates that an enriched macroscale continuum is needed for full consistency \cite{ramosExtensionHillMandel2017}, however \cite{larssonVariationallyConsistentComputational2010,ramosExtensionHillMandel2017,waseemModelReductionComputational2020} do utilize the standard continuum and that approach is also used here.

In the limit of temperature independent materials, a useful corollary of Eq.~\ref{eq:hom-energy-rate} is that the effective heat capacity is
\begin{equation}
    \bar{c} = \pdv{\bar{\epsilon}}{\bar{\theta}} = \pdv{}{\bar{\theta}}<\epsilon> = \pdv{}{\bar{\theta}}<\rho c_p \theta> = <\rho c_p>
\end{equation}
where we made use of the independence of \(\bar{\nabla}\bar{\theta}\) and \(\tilde{\theta}\) from the macro temperature.

Since the microscale fluctuations are not independent of the macroscale temperature gradient, the computation of the effective conductivity \(\vb{\kappa} = \pdv{\bar{\vb{q}}}{ \bar{\nabla}\bar{\theta}}\) requires the solution of time-dependent microscale problem in the general case or can be computed a-priori in the case of constant micro properties.

In \cite{bloomfieldMultiscaleWorkflowThermal2025}, we provide a method for computing the effective conductivity in the steady state setting. This pre-computation of the conductivity means that a single microscale computation can be used for the macroscale simulation. It also provides a natural way to use the results in a standard finite element solver such as ANSYS mechanical or Abaqus.

For the transient problem, there is less to be gained by computing effective conductivities and instead we directly utilize the homogenized heat flux in the time integration scheme. This is significantly more computationally efficient, it can be thought of as taking the action of the effective conductivity in the direction of the macro-solution, requiring only a single solve rather than the extraction of a full conductivity tensor that requires solves in each independent direction (3 for 3D).

\subsection{Time Integration}
The details of the time integration are slightly different for the macro- and micro-scales therefore, we provide the weak forms with each and briefly comment on their differences. In this initial implementation, we make use of the backward Euler time integration scheme for both scales since it is unconditionally stable. 

Applying the backward Euler method to the macroscale weak form given in Eq.~\ref{eq:weak-form} gives
\begin{equation}
    \int \delta \bar{\theta} \frac{\bar{\epsilon}^{n+1}-\bar{\epsilon}^{n}}{\Delta t}
    - \bar{\nabla} \delta\bar{\theta} \cdot \bar{\vb{q}}^{n+1} \dd{\Omega}+\int_{\Gamma_h} \delta \bar{\theta }\bar{\vb{q}}^{n+1}\cdot \vb{n} \dd{\Gamma} = 0
\end{equation}
which is an implicit equation noting that fluxes are needed at the \(n+1\) timestep. The values for $\bar{q}$ and $\bar{\epsilon}$ are obtained from the microscale solver at each integration point.

The microscale weak form makes use of two simplifications. Since homogeneous Dirichlet boundary conditions are used, there is no surface integral term to deal with. We also make use of the microscale constitutive equations, \(\dot{\epsilon} = \rho c_p \dot{\theta}\) and \(\vb{q} = \vb{\kappa} \cdot \nabla \theta \).
\begin{equation}
    \int \delta \tilde{\theta}\rho c_p \frac{\theta^{n+1}-\theta^{n}}{\Delta t}
    - \nabla \delta\tilde{\theta} \cdot \vb{\kappa} \cdot \nabla \theta^{n+1} \dd{\Omega} = 0.
\end{equation}
We note that due to the timescales involved in the microstructural model, subcycling may be desired to achieve the desired level of accuracy. Since the material properties are not a function of the temperature, the implicit equations are linear and can be solved with a matrix inversion at each scale.

The macroscale problem is integrated implicitly in time, while the microscale problems are synchronized at each macroscale time step and may be subcycled internally to resolve fast microstructural transients, with homogenized fluxes and energy rates returned to the macroscale at the end of each step.


\section{Test vehicle specification and property computations}
\subsection{SoC Layout}
\begin{figure}
    \centering
    \includegraphics[width=\linewidth]{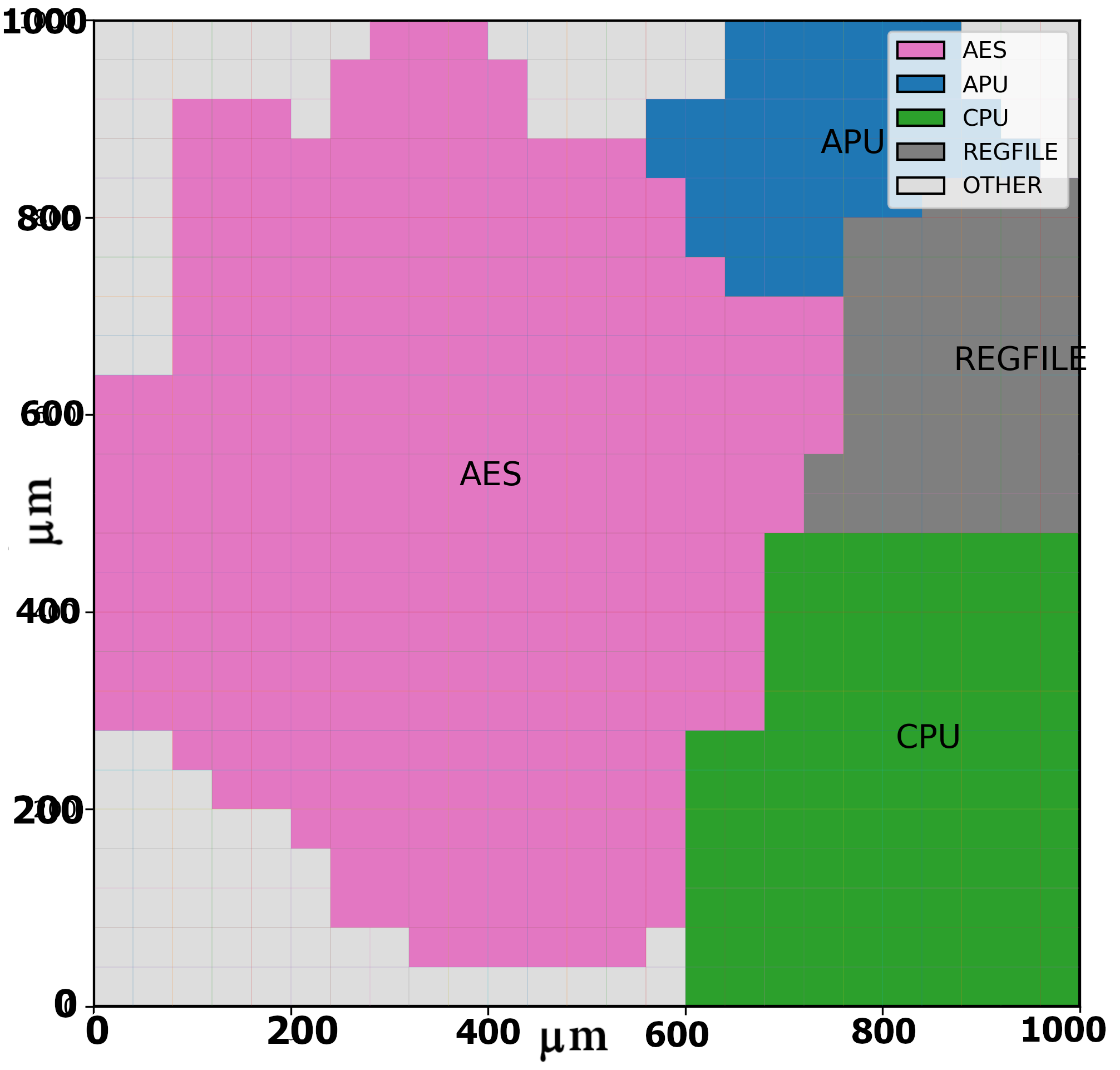}
    \caption{Tile-based dominance map of the LibreLane-generated Sky130A SoC test vehicle, showing the spatial distribution of CPU, cryptographic, peripheral, and memory logic. Each tile is assigned to the block with the highest standard-cell density in that region}
    \label{fig:tilemap}
\end{figure}

A 1$\times$1~mm$^2$~SoC-style model die was generated with LibreLane~\cite{librelane,fossi_foundation} to function as a non-proprietary test vehicle for the methodology. Based on the Sky130A PDK, it is composed of heterogeneous open-source compute (PicoRV32), crypto (AES), peripheral (APU), and memory array (2R1W) blocks. The design is a synthetic SoC-style test vehicle, not a production chip, created to produce realistic routed BEOL geometry and spatially varying power density for 3D heterogeneous integration and thermal modeling, with a layout of dominant cell densities as shown in Figure~\ref{fig:tilemap}. The BEOL for this die a FEOL/MOL-to-BEOL metal stack, comprising tungsten contacts and vias (LICON, MCON, V1–V4), aluminum interconnect layers M1 through M5 embedded in SiO$_2$ dielectric, and a top passivation layer.

\subsection{RVE Extraction Strategy}
All RVEs are extracted automatically from the GDS2 layout file and an XML-based description of the SkyWater 130 PDK, without intervention, using an in-house workflow as described in Ref.~\citen{bloomfieldMultiscaleWorkflowThermal2025}. BEOL layer elevations and thicknesses were defined based on the 130A PDK to enable construction of RVEs, while non-metal FEOL device layers were intentionally omitted. Representative materials properties were selected and are specified in Table~\ref{table:matprop}.

\begin{table}[h]
\centering
\caption{Thermal properties used in simulations}
\label{table:matprop}
\begin{tabular}{lccc}
\hline
Material & $\kappa$ (W/m·K) & $c_p$ (J/kg·K) & $\rho$ (g/cm$^3$) \\
\hline
Al        & 174.  & 900.  & 2.70  \\
W         & 62.   & 134. & 19.25 \\
SiO$_2$   & 1.07 & 1000.  & 2.20  \\
\hline
\end{tabular}
\end{table}

As discussed in the previous section, to solve a transient die-scale problem via a finite element method, the computational domain must be discretized (meshed) and the homogenized values for the volumetric heat capacity ($\langle \rho c_\mathrm{p} \rangle$) and thermal conductivity tensor $\bar{\kappa}$ are needed at every quadrature point in the discretization. In order to efficiently cache these properties for our SoC layout, we specify a regular grid over the die and compute these homogenized properties {\em a priori} at the grid points using the methods described in Section~\ref{sec:MTM}. Finally during macroscale transient simulations, we interpolate these properties onto the quadrature points once a discretization is provided. This approach is convenient for associating a map of the properties with the layout, independently of the mesh and thus independently of the larger 3D stack the die is to be included in. The trade off for this approach is that it does not allow for macroscale simulations requiring time-resolution of the subscale RVEs, such as when microscale inertia dominates. In those cases, explicit RVEs can be constructed and retained at the cost of increased computation and storage.

\begin{figure}
    \begin{subfigure}{0.4\linewidth}
    \centering
    \includegraphics[height=0.75\linewidth]{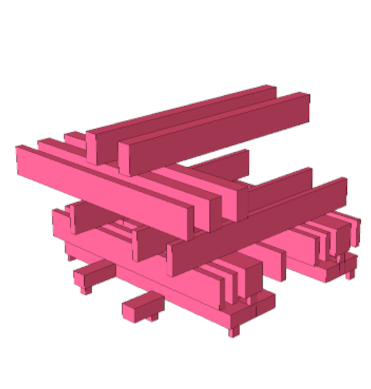}
    \caption{}
    \label{fig:5x5RVE}
    \end{subfigure}
    \begin{subfigure}{0.6\linewidth}
    \centering
    \includegraphics[height=0.5\linewidth]{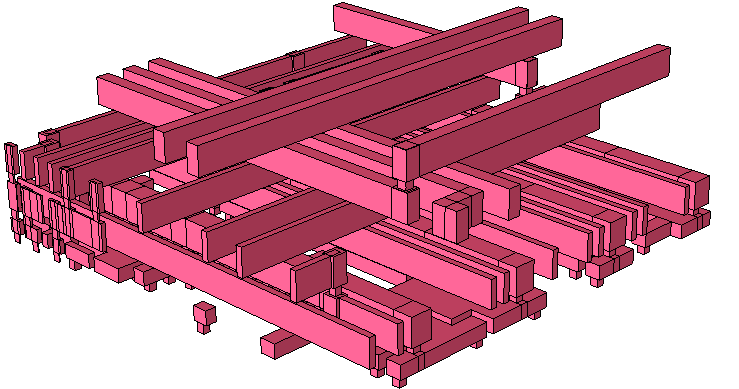}
    \caption{}
    \label{fig:10x10RVE}
    \end{subfigure}
    \caption{(a) Metal stack from a representative 5~\micron\ $\times$ 5~\micron\ RVE. Shown with the dielectric removed for visualization, this RVE has a volumetric heat capacity of 2.229~J/g.K and axis-aligned $\kappa$ values of $\bar{\kappa}_{xx}=12.5$, $\bar{\kappa}_{yy}=6.85$, and $\bar{\kappa}_{zz}=4.65$~W/m.K. (b) A similar 10~\micron\ $\times$~10~\micron\ RVE placed at the same point on the die. This RVE has a volumetric heat capacity of 2.227~J/g.K and axis-aligned $\kappa$ values of $\bar{\kappa}_{xx}=10.5$, $\bar{\kappa}_{yy}=4.97$, and $\bar{\kappa}_{zz}=3.18$~W/m.K.}
    \label{fig:rves}
\end{figure}

We constructed sets of RVEs of two different size for comparison and to help determine the sensitivity of the homogenized properties to RVE size for this layout. Both sets span the full thickness of the  BEOL described above (5.4~\micron). The in-plane extents of the RVEs are 5~\micron~$\times$~5~\micron\ and 10~\micron~$\times$~10~\micron{}. These two sets are distributed in uniform grids of 100x100 (10,000 RVEs) and 50x50 (2,500 RVEs) across the 1~mm~$\times$~1~mm die. Figure~\ref{fig:rves} shows examples of both sizes of RVE, extracted at the same point on the die, along with their homogenized properties. Comparison of results obtained from both RVE sizes at identical die locations demonstrates near-converged effective properties, justifying the use of the finer grid for dense spatial sampling. For each of the properties for the RVEs in Figure~\ref{fig:rves}, agreement is within 1\%.

\section{Results}
\subsection{Strongly Separated Regime}
In the strongly scale-separated regime, the microstructural response is assumed instantaneously equilibrate to the macrostructural state ($\bar{\nabla}\bar{\theta}, \bar{\theta})$. In this regime, the steady-state effective conductivity \(\bar{\kappa}^{ss}\) is sufficient to be used in the macroscale transient solution.

\begin{figure}[t]
  \centering
  \begin{subfigure}{0.49\linewidth}
    \centering
    \includegraphics[width=\linewidth]{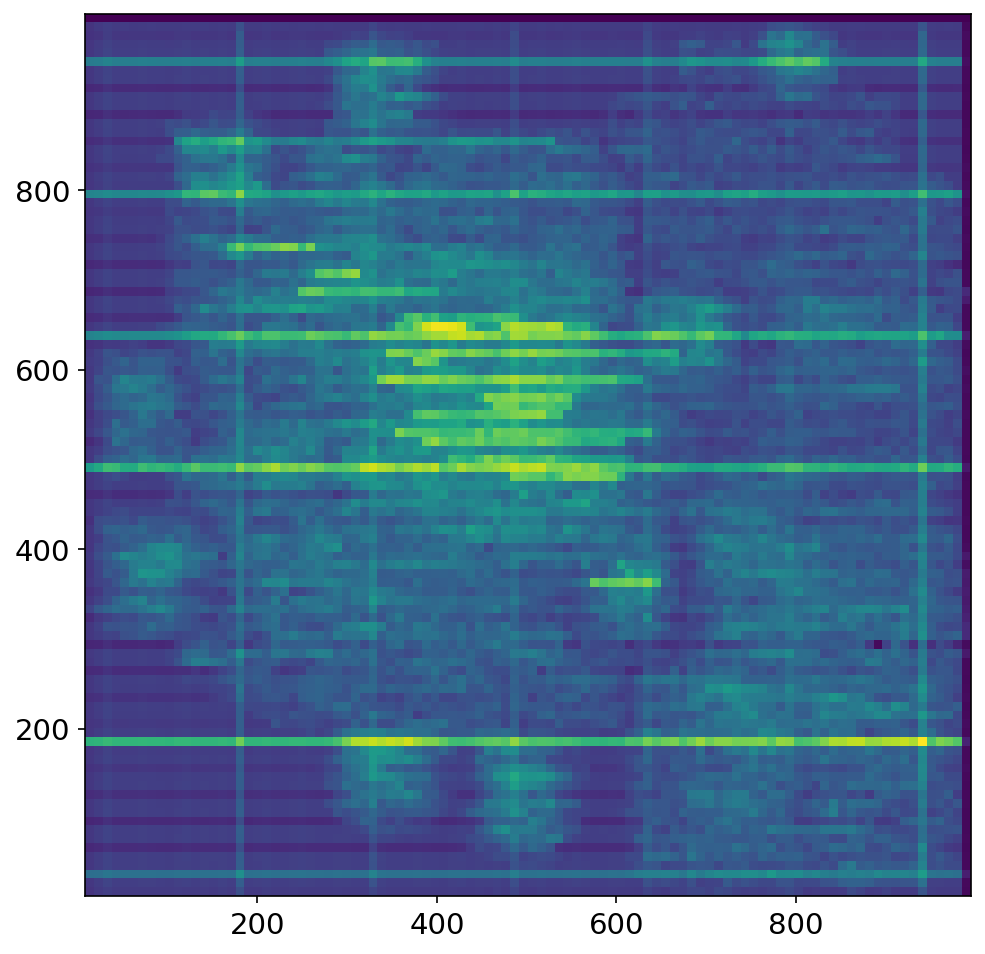}
    \caption{$\bar{\kappa}_{xx}^{ss}$ (5$\times$5~\micron)}
  \end{subfigure}\hfill
  \begin{subfigure}{0.49\linewidth}
    \centering
    \includegraphics[width=\linewidth]{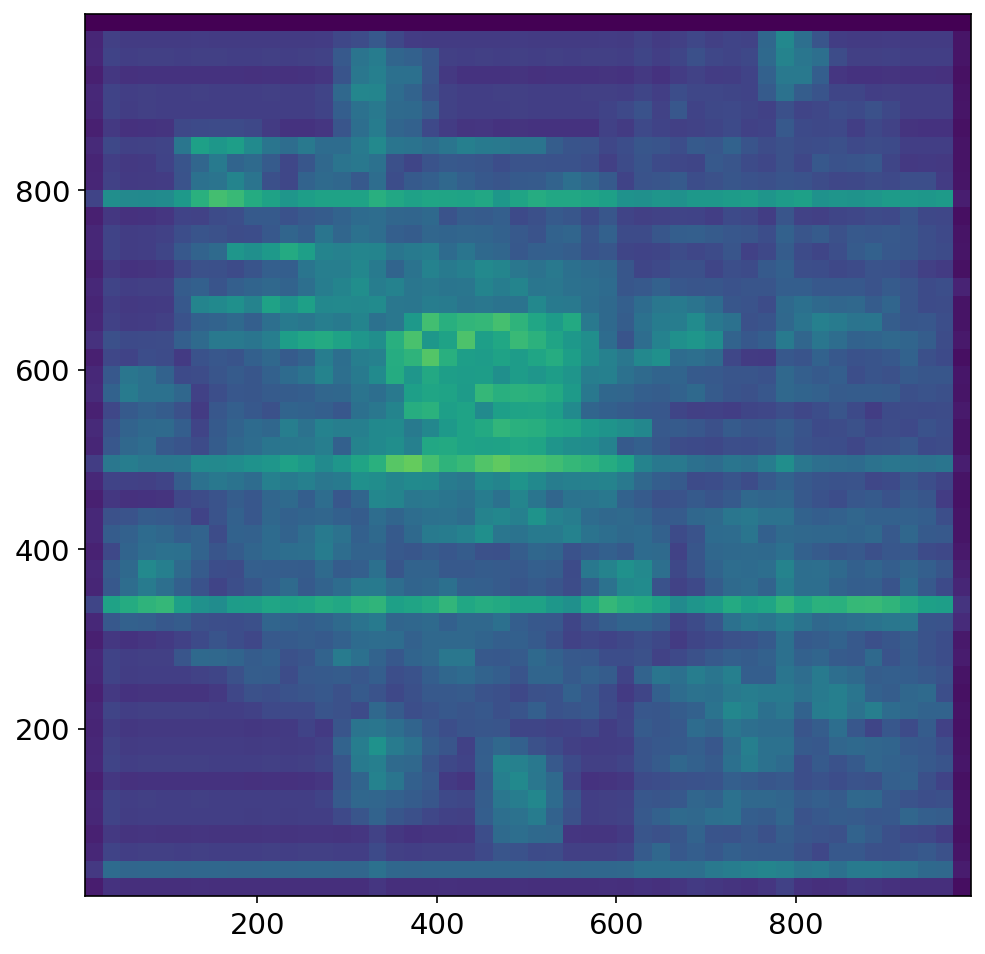}
    \caption{$\bar{\kappa}_{xx}^{ss}$ (10$\times$10~\micron)}
  \end{subfigure}
  \vspace{0.4em}
  \begin{subfigure}{0.49\linewidth}
    \centering
    \includegraphics[width=\linewidth]{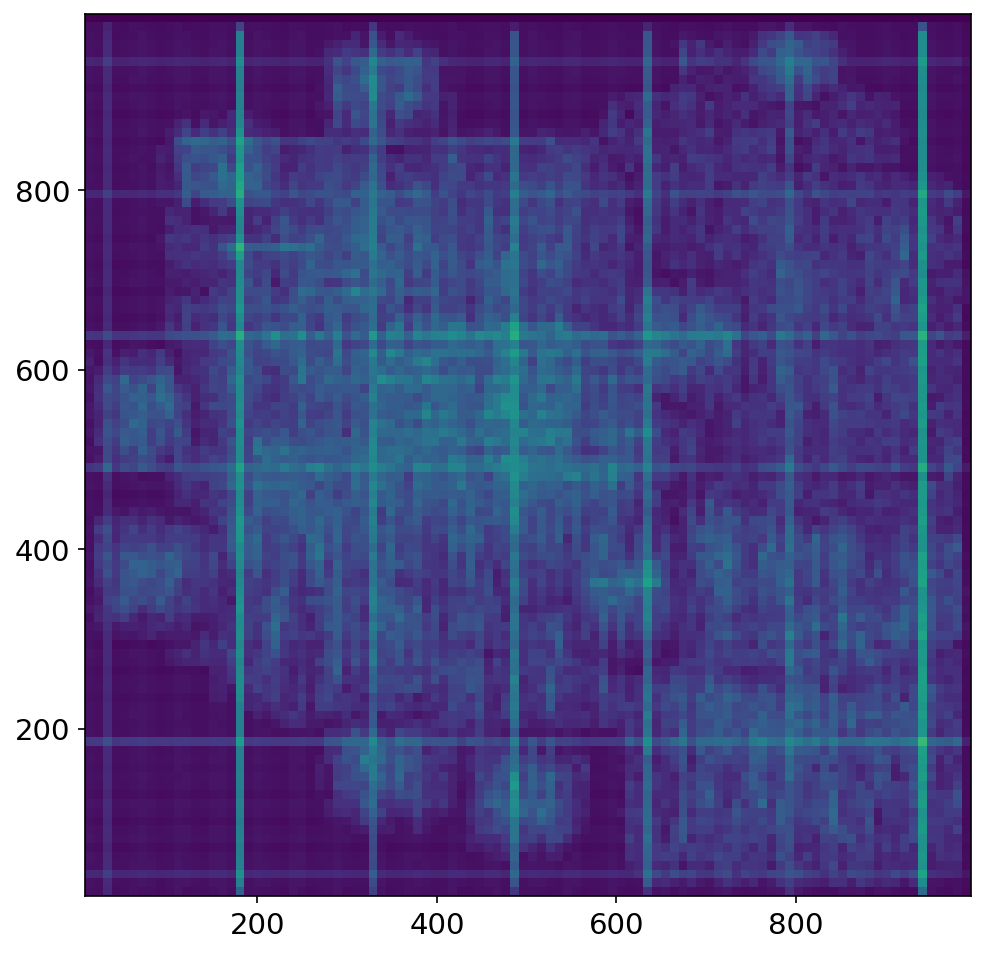}
    \caption{$\bar{\kappa}_{yy}^{ss}$ (5$\times$5~\micron)}
  \end{subfigure}\hfill
  \begin{subfigure}{0.49\linewidth}
    \centering
    \includegraphics[width=\linewidth]{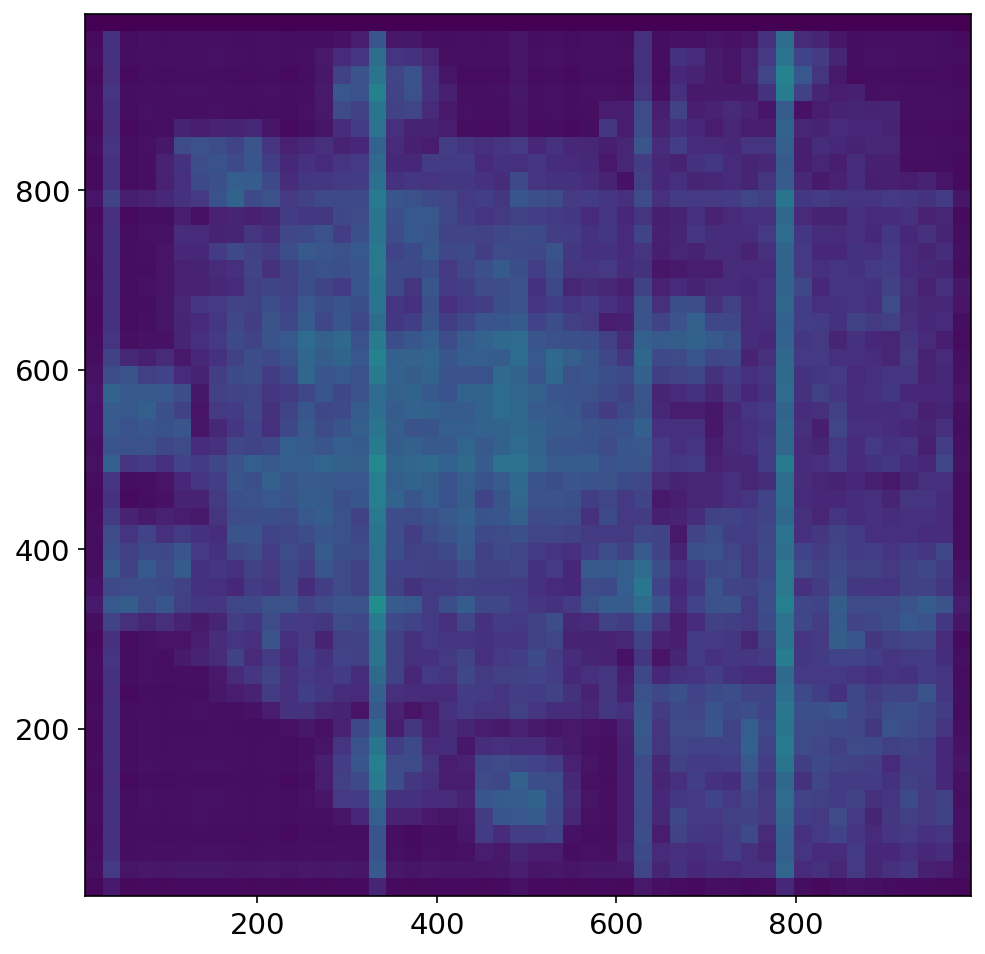}
    \caption{$\bar{\kappa}_{yy}^{ss}$ (10$\times$10~\micron)}
  \end{subfigure}
  \vspace{0.2em}
  \begin{subfigure}{0.49\linewidth}
    \centering
    \includegraphics[width=\linewidth]{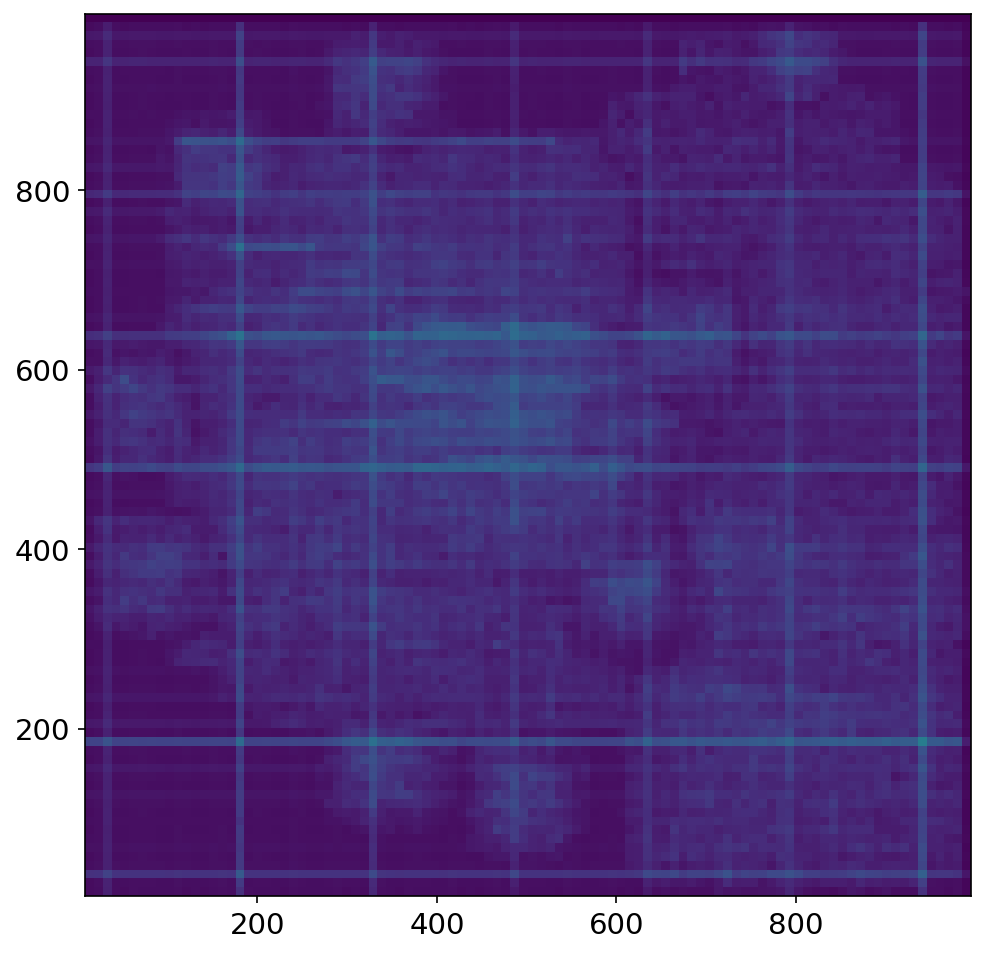}
    \caption{$\bar{\kappa}_{zz}^{ss}$ (5$\times$5~\micron)}
  \end{subfigure}\hfill
  \begin{subfigure}{0.49\linewidth}
    \centering
    \includegraphics[width=\linewidth]{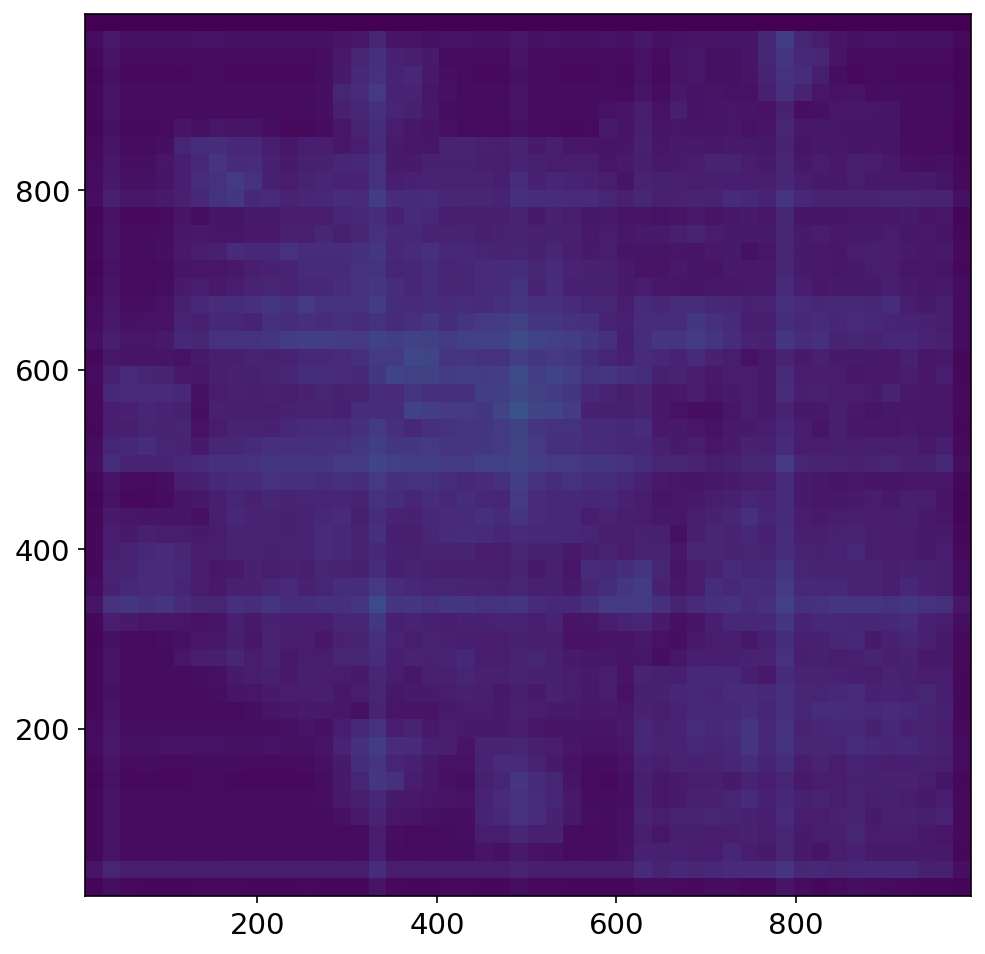}
    \caption{$\bar{\kappa}_{zz}^{ss}$ (10$\times$10~\micron)}
  \end{subfigure}
  \begin{subfigure}{0.99\linewidth}
    \includegraphics[width=\linewidth]{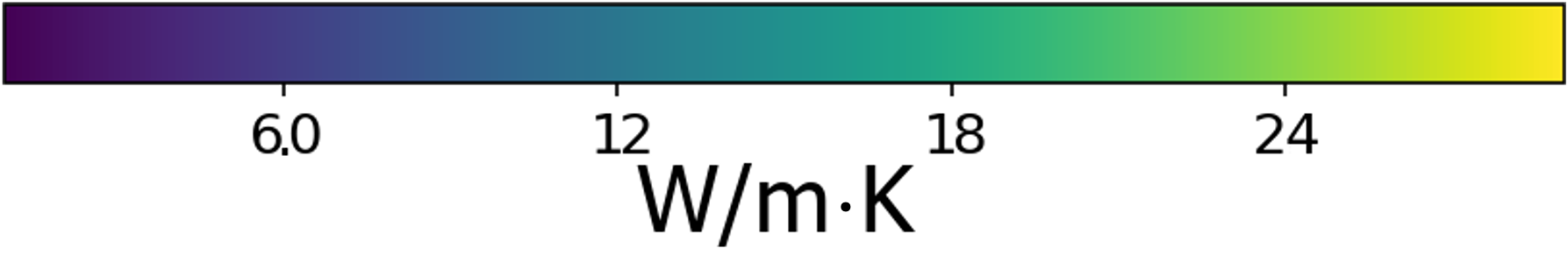}
  \end{subfigure}
  \caption{Spatial maps of the axis-aligned effective thermal conductivity components over the test vehicle die, computed from offline RVE analyses and reported on 5$\times$5~\micron\ (left column) and 10$\times$10~\micron\ (right column) sampling grids.}
  \label{fig:kappa_maps}
\end{figure}

\begin{figure}[t]
  \centering
  \begin{subfigure}{0.49\linewidth}
    \centering
    \includegraphics[width=\linewidth]{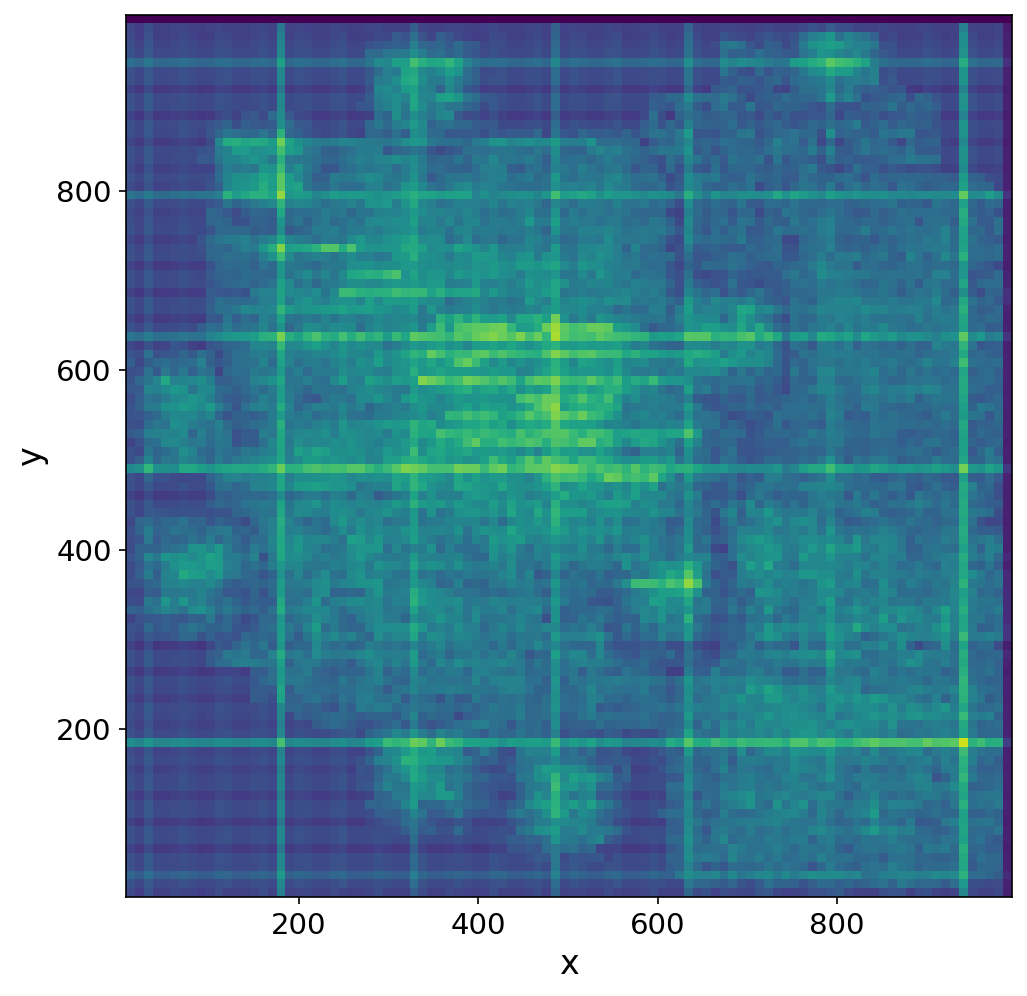}
    \caption{5$\times$5~\micron}
  \end{subfigure}\hfill
  \begin{subfigure}{0.49\linewidth}
    \centering
    \includegraphics[width=\linewidth]{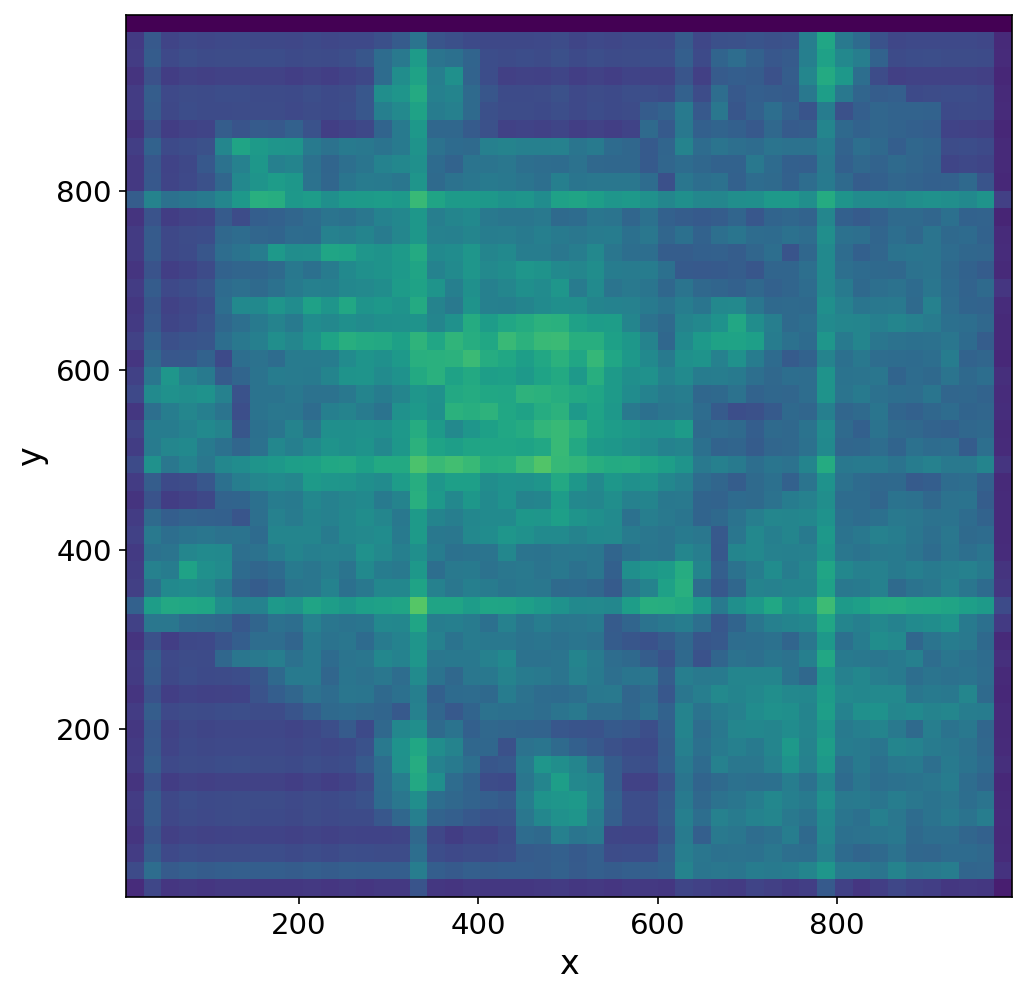}
    \caption{10$\times$10~\micron}
  \end{subfigure}
  \begin{subfigure}{0.99\linewidth}
    \includegraphics[width=\linewidth]{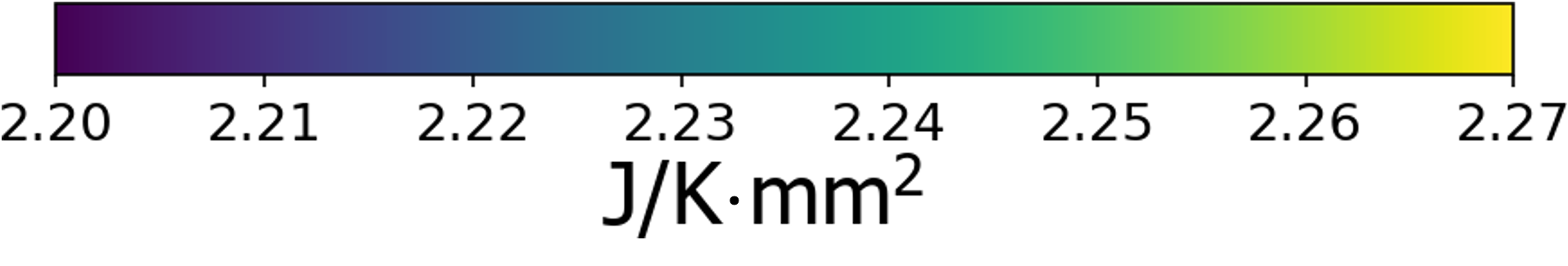}
  \end{subfigure}  \caption{Spatial maps of the effective volumetric heat capacity $\rho \mathrm{c}_p$ over the test vehicle die, computed from offline RVE analyses and reported on 5$\times$5~\micron\ and 10$\times$10~\micron\ sampling grids. These fields are tabulated on the sampling grid and interpolated for use in the macroscale transient simulation.}
  \label{fig:rhocp_maps}
\end{figure}

The spatial maps of the principal thermal conductivity tensor components and the volumetric heat capacities are shown in Figures~\ref{fig:kappa_maps} and ~\ref{fig:rhocp_maps}. Each figure shows both the 5~\micron\ $\times$ 5~\micron\ RVE result on the left and the 10~\micron\ $\times$ 10~\micron\ RVE result on the right, with a common scale for all six figures. The result for both RVE sizes show very similar spatial distributions, and the in-plane conductivities are greater in the ``x'' direction, with the out-of-plane conductivities being consistently lower. 

With both RVE sizes, the actual structural sampling of BEOL structures is 25\%, with the $100\times100$ grid representing 10,000 RVEs of $2.5\times 10^{-5}$~mm$^2$ each, for a total of 0.25~mm$^2$ over the 1~mm$^2$ die. The total sampled area for the grid of 2,500 10~\micron~$\times~10$~\micron\ RVEs has the same area. However, geometric model construction takes considerably more memory and is more prone to failure for higher-complexity RVEs, \textit{i.e.}, when tiles contain many disconnected components. Thus, it is convenient to use the smallest RVE possible.

It is notable that the long thin power delivery straps that run horizontally (the ``x'' direction), are not well captured by this approach alone, particularly in light of their alignment with our grids. These pairs of 1.6~\micron\ aluminum lines in Metal 5 cross the entire chip and can act as significant pathways for heat conduction in bands across the die. However, although these types of structures are very high aspect ratio, they are not complex in geometry and can be directly and efficiently represented in any macroscale calculation, most likely with a highly refined or anisotropic mesh in those regions.

\subsection{Weak Scale Separation}
When scale separation is violated, the effective conductivity can be written as
\begin{equation}
     \bar{\vb*{\kappa}} = <\vb*{\kappa}  \cdot (\vb*{\mathrm{I}}+ \nabla \vb{w}^{\nabla \theta}) + \frac{\rho c_p}{\Delta t} (\vb{x}-\bar{\vb{x}})\otimes
   ((\vb{x}-\bar{\vb{x}})+\vb{w}^{\nabla \theta})>
   \label{eq:kappa-transient}
\end{equation}
where, \(\vb{w}^{\nabla \theta}\) is sensitivity of the microscale fluctuations given by to the macroscale temperature gradient that can be solved by an auxiliary problem given by
\begin{equation}
    \pdv{R}{\theta} w^{\bar{\nabla}\bar{\theta}} =  -\int_{\Omega}
     \delta \tilde{\theta} \frac{\rho c_p}{\Delta t} (\vb{x}-\bar{\vb{x}})
    -\nabla \delta \tilde{\theta} \cdot \vb*{\kappa} \dd{\Omega}.
\end{equation}
In the fully linearized problem, there are also two vector operators that provide a coupling between the macroscale temperature and macroscale temperature gradients. These two terms are not compatible with commercial FEM codes, and have not been addressed here. A more full exposition of the role of these coupling parameters will be addressed in future work.

\begin{figure}
  \centering
  \begin{subfigure}{0.32\linewidth}
    \centering
    \includegraphics[width=\linewidth]{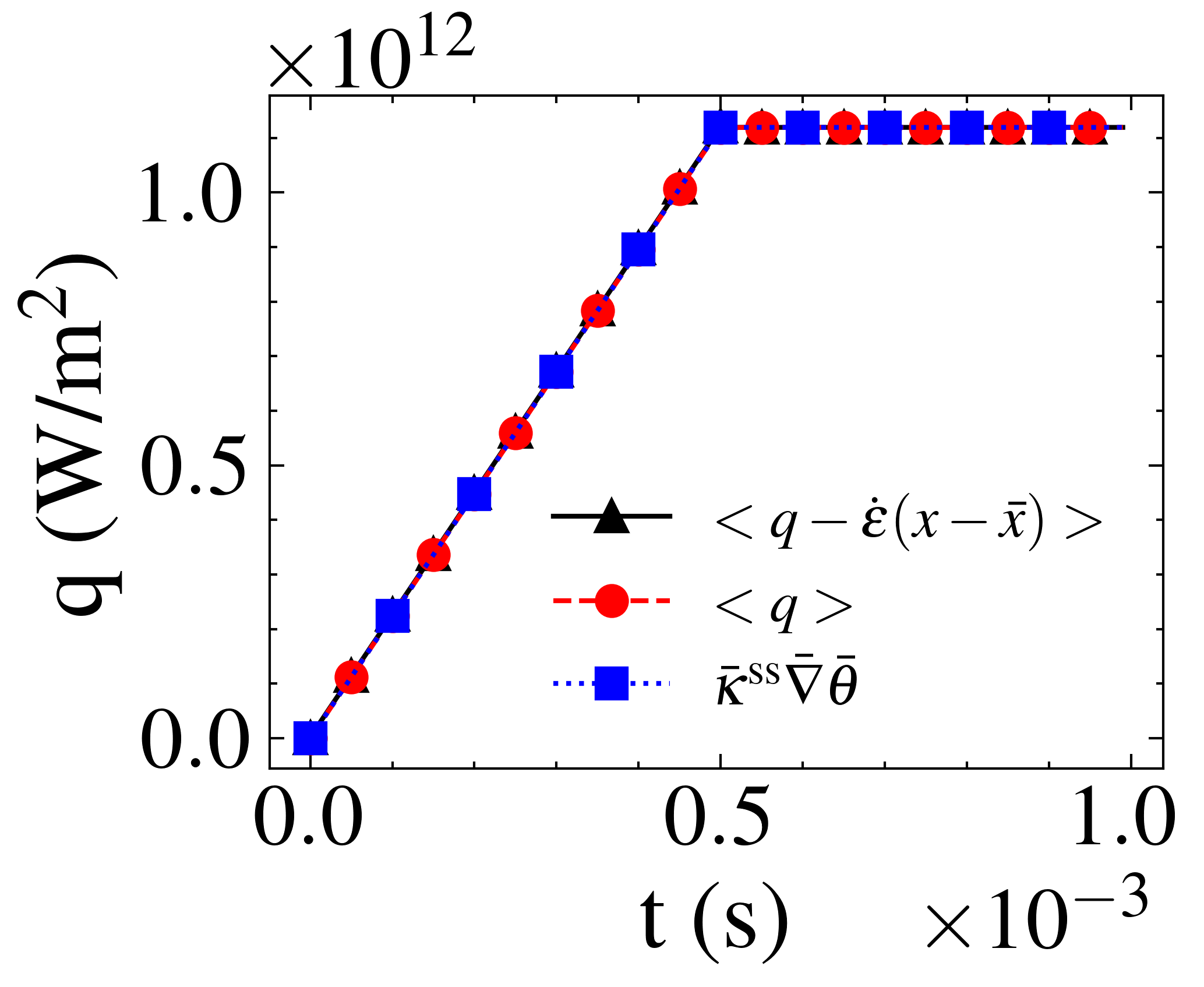}
    \caption{$t_L = 5\times 10^{-4}s$}
  \end{subfigure}\hfill
  \begin{subfigure}{0.32\linewidth}
    \centering
    \includegraphics[width=\linewidth]{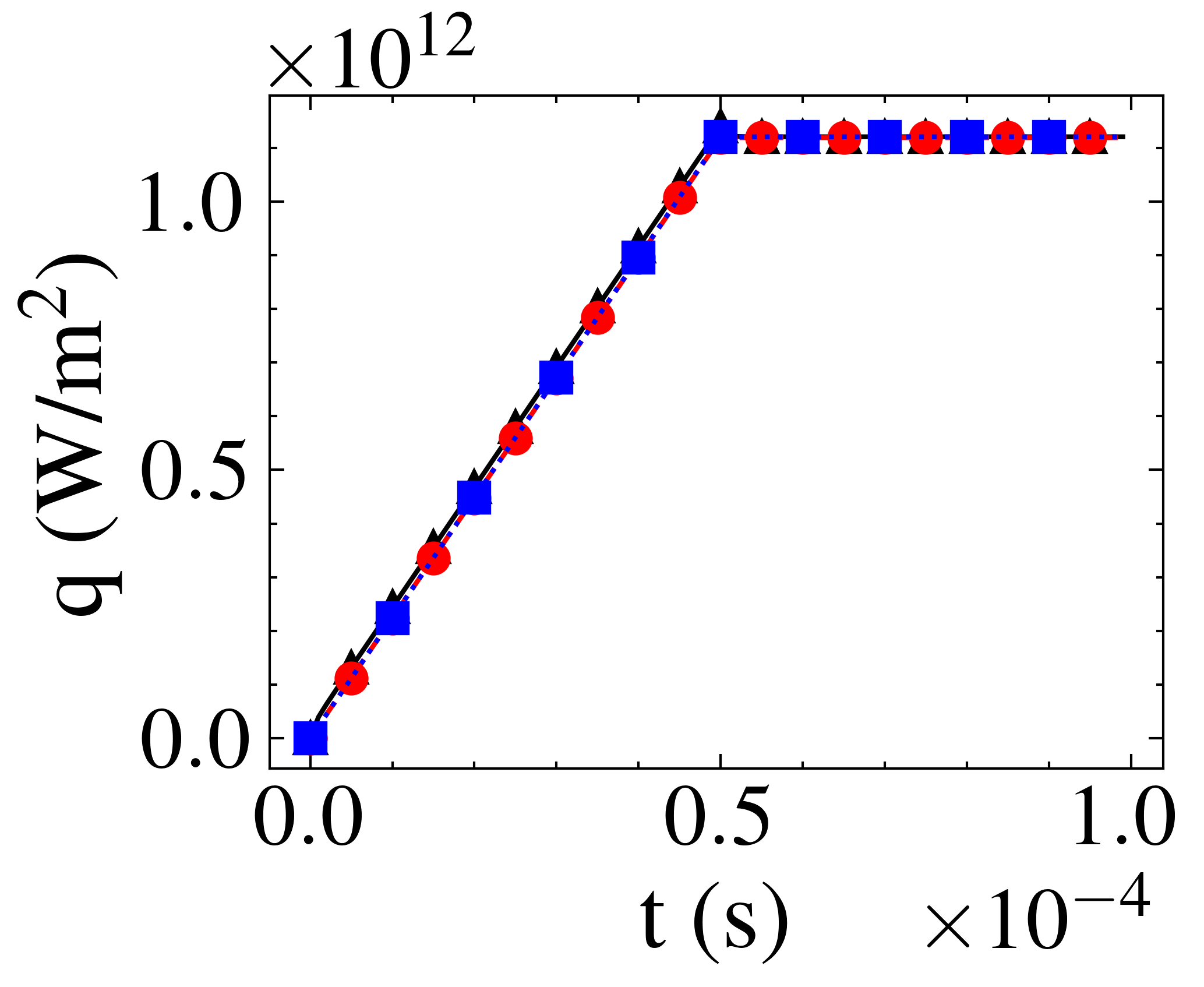}
    \caption{$t_L = 5\times 10^{-5}s$}
  \end{subfigure}
  \begin{subfigure}{0.32\linewidth}
    \centering
    \includegraphics[width=\linewidth]{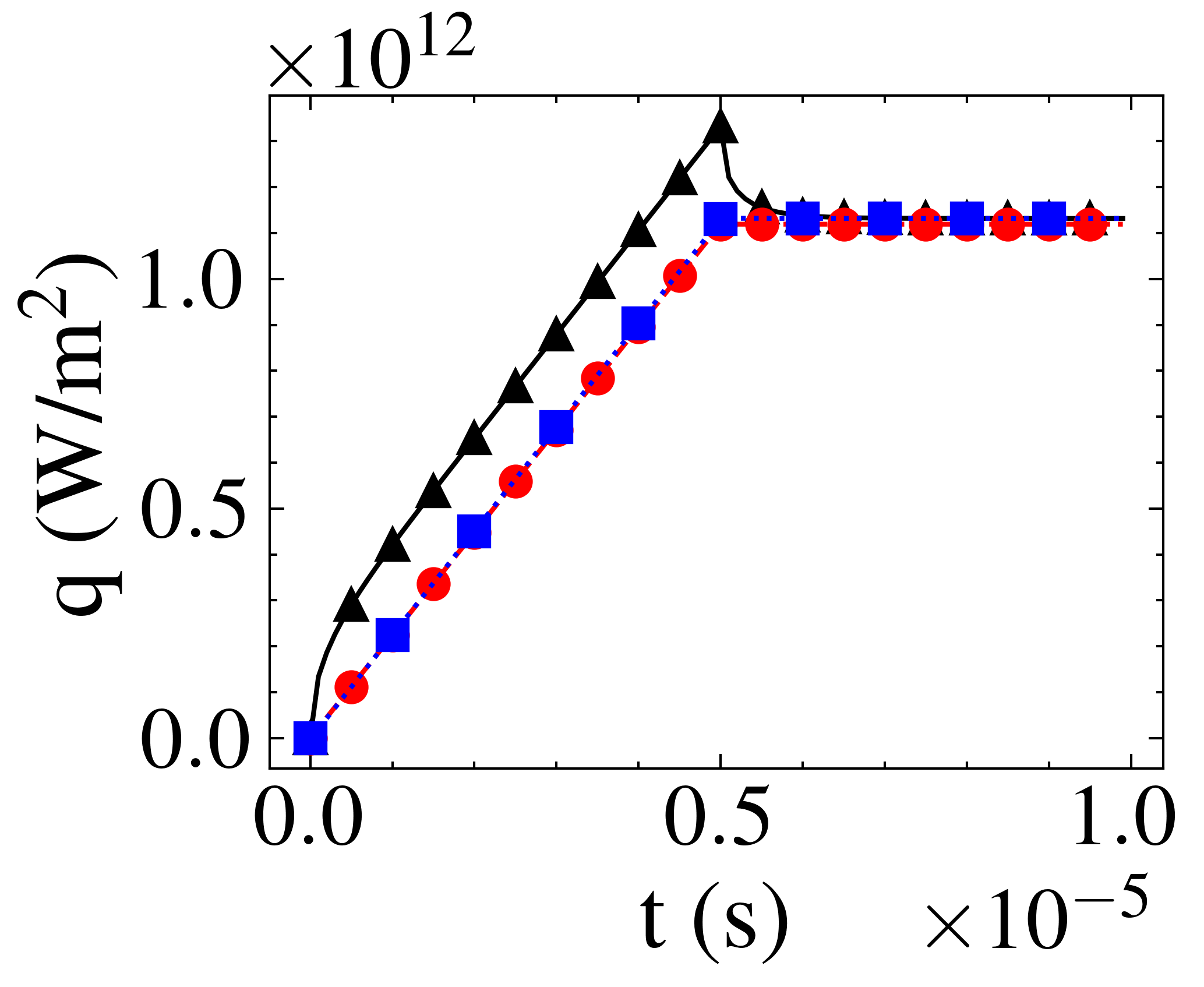}
    \caption{$t_L = 5\times 10^{-6}s$}
  \end{subfigure}
    
    \caption{Impact of transient contributions to the homogenized conductivity for a single 10$\times$10~\micron\ RVE subjected to temperature gradient controlled boundary conditions with the maximum temperature difference in the through-thickness direction reaching 1K after a linear ramp period of $t_L$. The three curves compare the variationally consistent transient homogenized flux, the plain flux average (without the inertia correction), and the steady-state prediction, illustrating when microscale thermal inertia becomes important as the loading rate increases.}
    \label{fig:single-rve}
\end{figure}

To quantify the impact of the transient terms on the homogenized heat flux Eq.~\ref{eq:kappa-transient} we plot the homogenized heat flux computed with and without the transient terms in Fig.~\ref{fig:single-rve} under three different loading rates. The red and blue curves provide two common reference predictions (naive flux average and steady-state constitutive response), while the black curve is the variationally consistent transient homogenized flux. A temperature gradient boundary condition is applied to the RVE with a zero gradient across the in-plane directions and a gradient that enforces a maximum of a 1K temperature difference across the through-thickness direction. The gradient is applied in a linear ramp fashion for \(t_L = 5\times10^{-4}s\), \(t_L = 5\times10^{-5}s\), and \(t_L = 5\times10^{-6}s\) and held at a constant value thereafter.

For a slow loading rate, relative to the chosen materials and microstructure, shown in Fig.~\ref{fig:single-rve}a, the transient part of the homogenized heat flux has negligible impact on the evolution of the RVE. When the loading rate is increased Fig.~\ref{fig:single-rve}b and Fig.~\ref{fig:single-rve}c there is an increase in the amount of heat flux needed to equilibrate to the enforced temperature gradients. This increase is the impact of the RVE thermal inertia and demonstrates that as the macroscale loading rate increases, it becomes more important to include the microstructural transient effects in the homogenized fluxes.

\begin{figure}
    \centering
    \begin{subfigure}{0.32\linewidth}
    \centering
    \includegraphics[width=\linewidth]{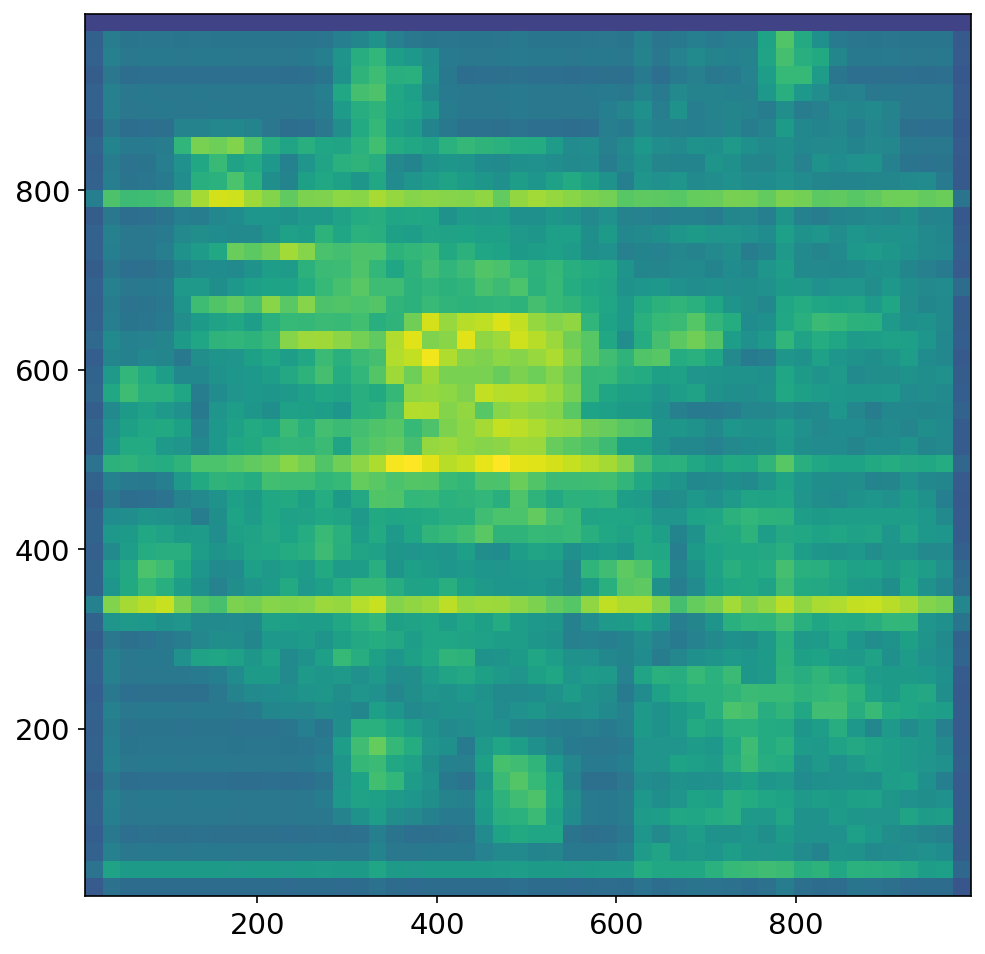}
    \caption{$\bar{\kappa}_{xx}$}
    \end{subfigure}
    \begin{subfigure}{0.32\linewidth}
    \centering
    \includegraphics[width=\linewidth]{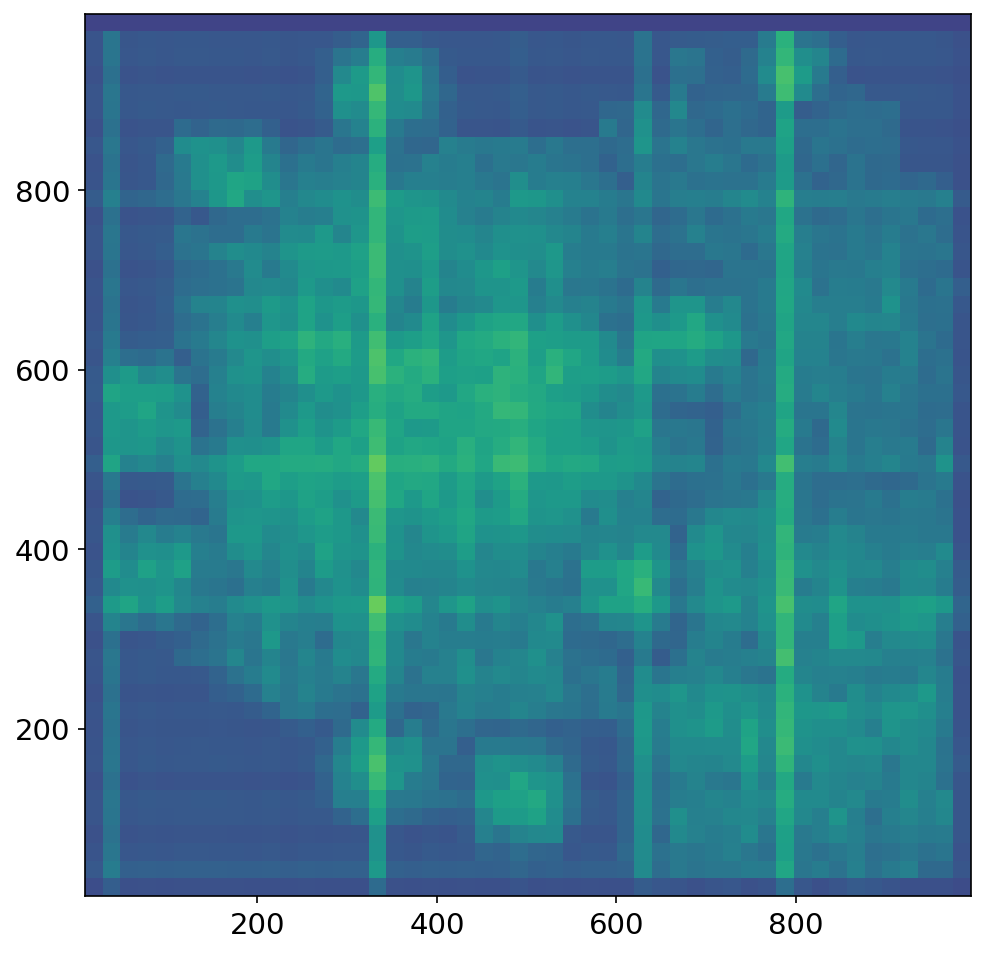}
    \caption{$\bar{\kappa}_{yy}$}
    \end{subfigure}
    \begin{subfigure}{0.32\linewidth}
    \centering
    \includegraphics[width=\linewidth]{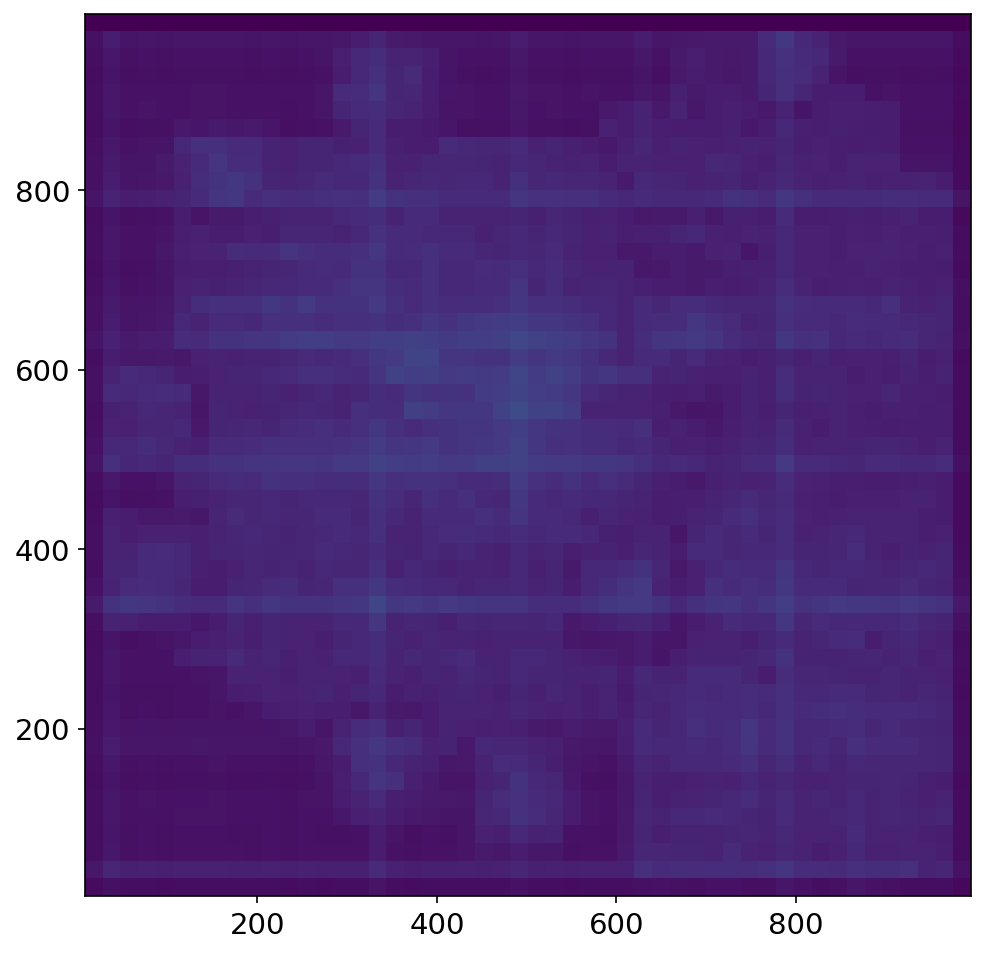}
    \caption{$\bar{\kappa}_{zz}$}
    \end{subfigure}
    \vspace{0.5em}
    \begin{subfigure}{\linewidth}
    \centering
    \includegraphics[width=0.85\linewidth]{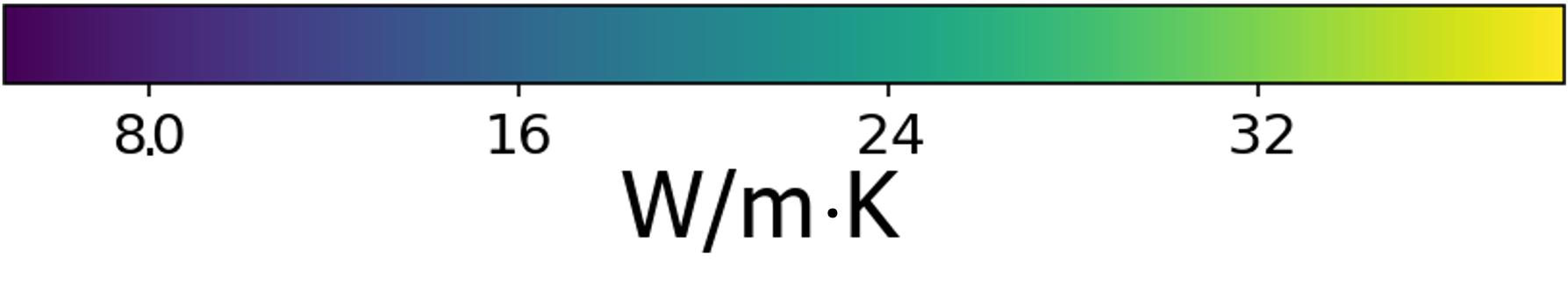}
    \end{subfigure}
    \caption{
        Spatial maps of the axis-aligned effective transient thermal conductivity components over the test vehicle die, computed from offline RVE analyses and reported for a 10$\times$10~\micron\ RVE size and a 50$\times$50 sampling grid.
    }
    \label{fig:transient-effective-conductivity-map}
\end{figure}

In the case of a constant integration timestep and temperature independent microscale material properties, the effective conductivity given in Eq.~\ref{eq:kappa-transient} can be pre-computed in an offline step. Figure~\ref{fig:transient-effective-conductivity-map} shows the property maps for a timestep $\Delta t = 0.001~\text{s}$. 

\begin{figure}[t]
  \centering
  \begin{subfigure}{0.49\linewidth}
    \centering
    \includegraphics[width=\linewidth]{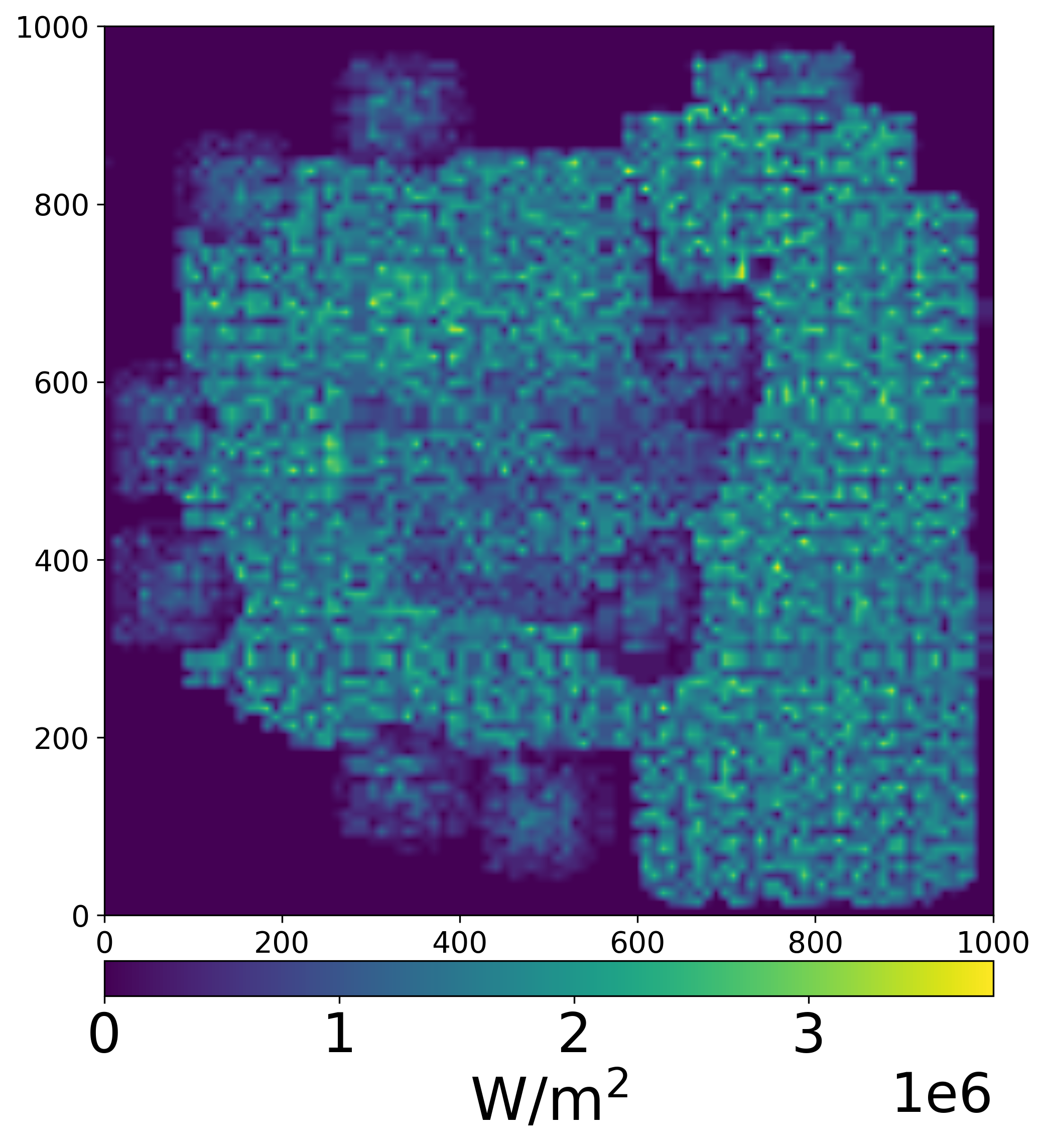}
    \caption{Applied heat flux map.}
    \label{fig:flux_map}
  \end{subfigure}\hfill
  \begin{subfigure}{0.49\linewidth}
    \centering
    \includegraphics[width=\linewidth]{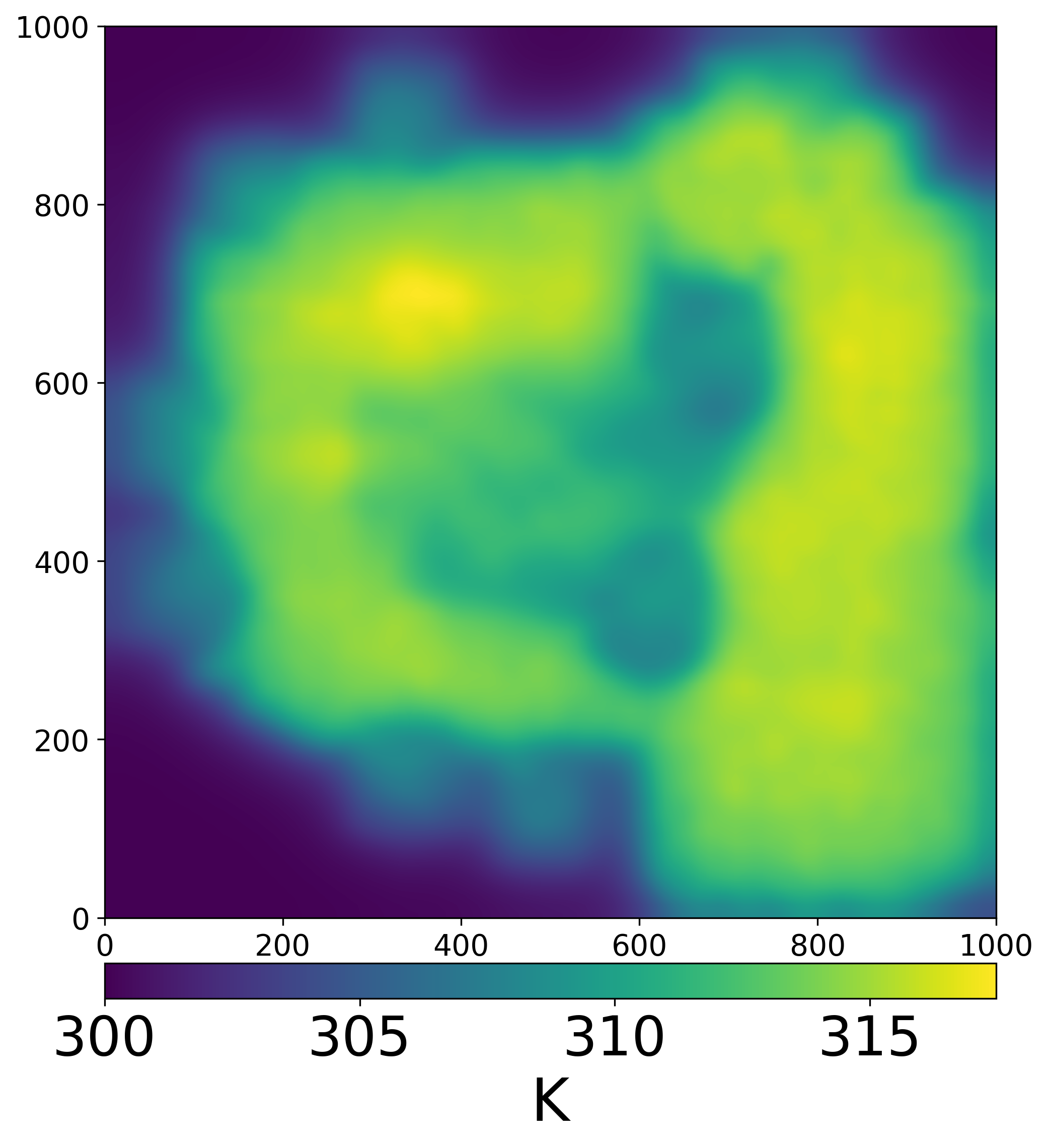}
    \caption{Temperature field.}
    \label{fig:temp_map}
  \end{subfigure}
  \caption{Macroscale demonstration: (a) applied surface heat-flux map and (b) resulting die-scale final temperature field.}
  \label{fig:flux_and_temp}
\end{figure}

The macroscale demonstration uses a spatially varying surface heat-flux map applied on the bottom surface of the die as a Neumann boundary condition.
The applied flux distribution and the resulting temperature field at the final time are shown in Fig.~\ref{fig:flux_and_temp}. After the response reaches a quasi-steady state, the macroscale hotspot and average temperatures are $T_{\max}=318.72~\text{K}$ and $T_{\mathrm{avg}}=310.44~\text{K}$.

\subsection{Validation against a fully resolved 50$\times$50~\micron\ BEOL model}

\begin{figure}[htb]
  \centering
  \includegraphics[width=\linewidth]{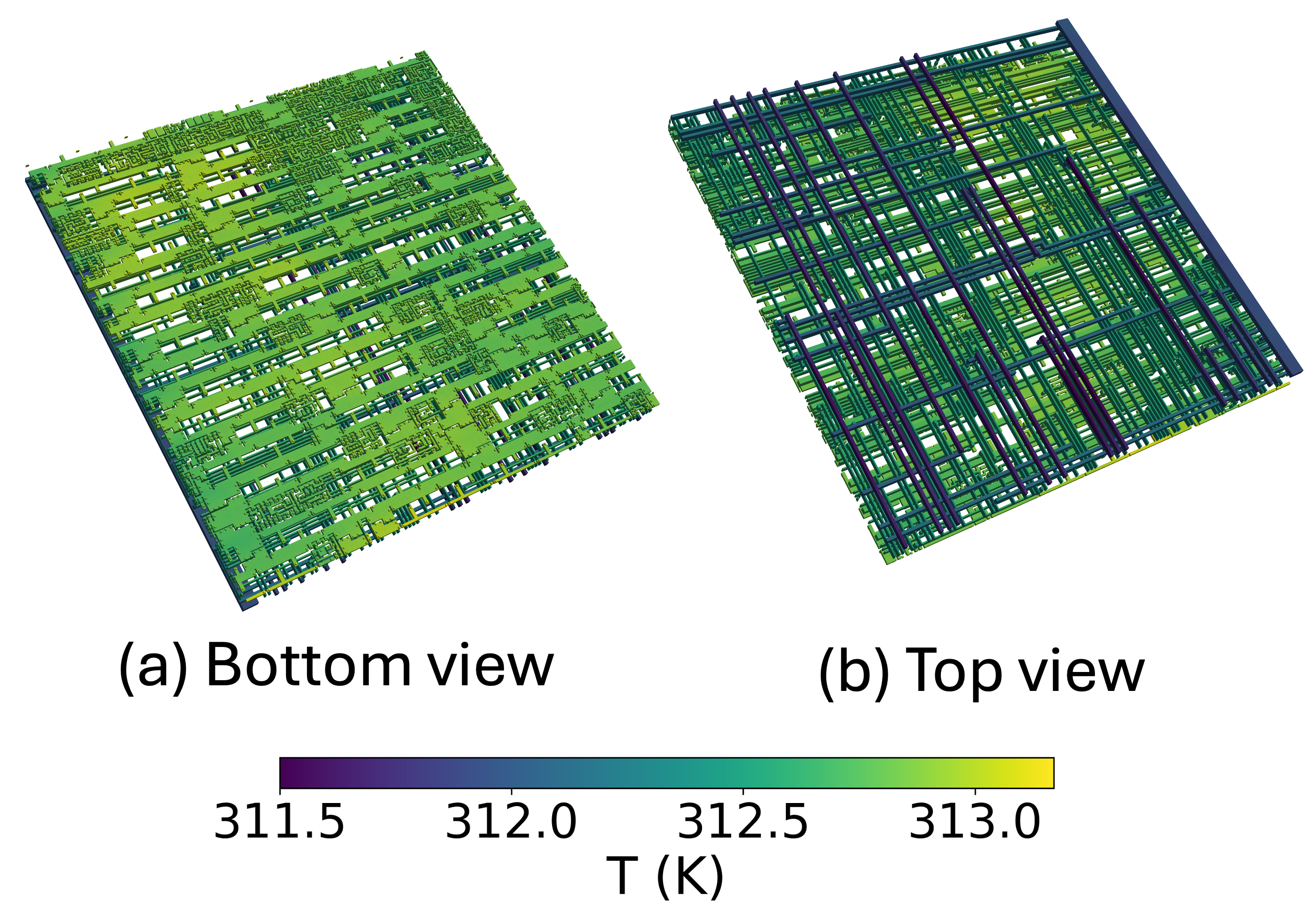}
  \caption{Fully resolved 50$\times$50~\micron\ BEOL reference geometry used
  for validation: (a) bottom view and (b) top view. The dielectric is
  removed for visualization clarity.}
  \label{fig:beol50_resolved_geometry}
\end{figure}

\begin{figure}[t]
  \centering
  \includegraphics[width=\linewidth]{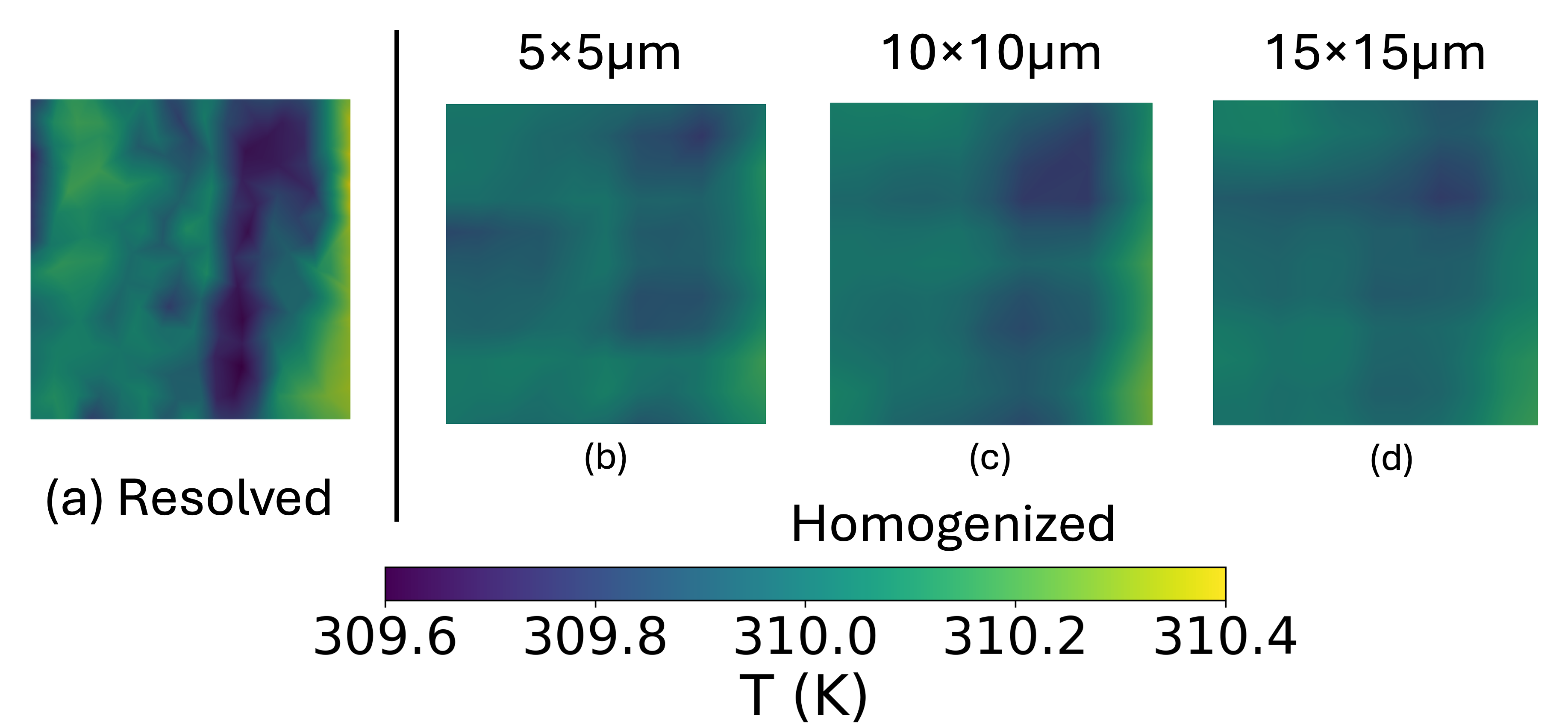}
  \caption{Top-surface temperature field comparison for the steady-state
  validation case: (a) fully resolved model and homogenized models using
  (b) 5$\times$5~\micron, (c) 10$\times$10~\micron, and
  (d) 15$\times$15~\micron\ RVEs.}
  \label{fig:beol50_validation_maps}
\end{figure}

To validate the homogenized model, we compare it against a fully resolved
steady-state simulation of an explicit 50$\times$50~\micron\ BEOL geometry
containing Al, W, and SiO$_2$ regions. Both models use the same boundary
conditions: a uniform Neumann heat flux of $10^{6}$~W/m$^2$ is applied on
the bottom surface, and a Robin condition is imposed on the top surface
with $T_{\mathrm{amb}}=300$~K and $h=10^{5}$~W/(m$^2\cdot$K). We use the
steady-state response for validation because it isolates the conductivity-
driven behavior represented by the homogenized conductivity maps and avoids
additional sensitivity to time discretization and initial conditions.

Figure~\ref{fig:beol50_resolved_geometry} shows the fully resolved
reference geometry from the bottom and top views, with the dielectric
removed for visualization clarity. This highlights the geometric
complexity of the explicit BEOL structure. Figure~\ref{fig:beol50_validation_maps}
compares the top-surface temperature field from the fully resolved model
with homogenized predictions obtained using 5$\times$5, 10$\times$10, and
15$\times$15~\micron\ RVEs. Although the homogenized fields do not reproduce
the fine-scale variations associated with the explicitly resolved
interconnect geometry, they capture the dominant large-scale temperature
pattern observed in the resolved solution (including the location and shape
of the main hot/cool regions), with modest changes as the RVE size is
increased.

\begin{table}[htb]
\centering
\caption{Validation against the fully resolved
50$\times$50~\micron\ BEOL model.}
\label{tab:beol50_ss_validation}
\begin{tabular}{lccc}
\hline
Model & $T_{\mathrm{top,avg}}$ (K) & $T_{\mathrm{bot,avg}}$ (K) & Error (\%) \\
\hline
Fully resolved & 310.00 & 312.97 & - \\
Homog., 5$\times$5~\micron\ RVE   & 310.00 & 311.37 & 0.51 \\
Homog., 10$\times$10~\micron\ RVE & 310.00 & 311.56 & 0.45 \\
Homog., 15$\times$15~\micron\ RVE & 310.00 & 311.79 & 0.38 \\
\hline
\end{tabular}
\end{table}

For quantitative comparison, Table~\ref{tab:beol50_ss_validation} reports
the average top and bottom surface temperatures and the percent error in
$T_{\mathrm{bot,avg}}$ relative to the fully resolved result.
All cases give $T_{\mathrm{top,avg}}\approx310$~K, and the homogenized
predictions for $T_{\mathrm{bot,avg}}$ are within 0.38--0.51\% of the fully
resolved value. Thus, the homogenized model captures the main large-scale
temperature pattern and shows a systematic improvement in the bottom-surface
average temperature as the RVE size increases.
r

\section{Conclusions and Limitations}
In this paper, we have demonstrated a transient multiscale simulation workflow for BEOL structures that are automatically constructed from GDSII/OASIS files. This method does not rely upon 1D heat flow assumptions that are common in other transient methodologies for microelectronic devices.

Under the assumptions of temperature independent microstructural constitutive properties and constant time integration timestep, a constant effective conductivity and heat capacity are obtained. These properties can be directly used in commercial transient FEM-based thermal solvers as standard, linear constitutive models. Spatial maps of these properties are provided under the assumption of strong and weak times-scale separation.

Although these property maps can be effectively used in the commercial FEM codes, they do not account for the full linearized structure of the transient FEM problem. Instead, two additional vector constants appear that couple the macroscale temperature and temperature gradients. We plan to quantify the impact of these coupling terms in our future work.

\bibliography{references}
\end{document}